\documentclass[a4paper]{article}

\usepackage{INTERSPEECH2021}

\RequirePackage{booktabs}
\usepackage{amsmath,graphicx}

\usepackage{xcolor}
\usepackage{multirow}

\usepackage{float}

\title{Bootstrap an End-to-end ASR System by Multilingual Training, \\ Transfer Learning, Text-to-text Mapping and Synthetic Audio}
\name{Manuel Giollo*$^1$, Deniz Gunceler*$^2$, Yulan Liu*$^3$, Daniel Willett$^2$ \thanks{*These authors have contributed equally.}}
\address{
    $^1$Alexa Speech, Amazon Development Centre, Turin, Italy \\
    $^2$Alexa Speech, Amazon Development Center GmbH, Aachen, Germany \\
    $^3$Alexa Speech, Amazon Development Center, Cambridge, UK
}

\email{\{mgiollo,denizg,lyulan,dawillet\}@amazon.com}

\begin{document}

\maketitle

\begin{abstract}
Bootstrapping speech recognition on limited data resources
has been an area of active research for long. 
The recent transition to all-neural models and end-to-end (E2E) training brought along 
particular challenges as these models are known to be data hungry,
but also came with opportunities around language-agnostic representations
derived from multilingual data as well as
shared word-piece output representations across languages that share script and roots.
We investigate here the effectiveness of different strategies
to bootstrap an RNN-Transducer (RNN-T) based automatic speech recognition (ASR) 
system in the low resource regime,
while exploiting the abundant resources available in other languages
as well as the synthetic audio from a text-to-speech (TTS) engine.
Our experiments demonstrate that transfer learning from a multilingual model,
using a post-ASR text-to-text mapping
and synthetic audio deliver additive improvements,
allowing us to bootstrap a model for a new language
with a fraction of the data that would otherwise be needed.
The best system achieved a 46\% relative word error rate (WER) reduction compared to the monolingual baseline,
among which 25\% relative WER improvement is attributed to the post-ASR text-to-text mappings
and the TTS synthetic data.

\end{abstract}


\section{Introduction}
An automatic speech recognition (ASR) system is an essential component of a voice 
assistant.
Building an ASR system which meets the accuracy bar for commercial
products can take thousands of hours 
of labelled in-domain audio for each target language.
Traditionally, an ASR system is first built with 
a relatively small amount of available labelled audio, and this process is 
referred to as \textit{ASR bootstrapping}.
Such an ASR system is used in beta instalments to gradually collect more 
in-domain target language data which in return is used to incrementally improve
the ASR system.

Recently, end-to-end (E2E) ASR models have emerged as the state-of-the-art for
speech recognition 
\cite{Chan2016LAS,Bahdanau2016,rnnt}.
However, very few studies have thoroughly compared or combined the
bootstrapping strategies for E2E models with no or limited labelled in-domain data 
(e.g. $<$200h) but still achieving the high recognition accuracy on realistic audio 
traffic that a voice assistant product needs.
This work investigates an ASR system based on recurrent neural network transducer 
(RNN-T) \cite{rnnt,graves2012sequence}
at such a low data limit during the bootstrapping phase of 
creating a new ASR system for a previously not covered language.

The ASR bootstrapping scenario in this work assumes that a large amount of 
transcribed audio is available in English, German, Spanish and French.
The goal is to bootstrap an ASR system 
for a similar voice assistant application in Italian with as little data as possible,
especially with as little labelled audio recordings as possible as they are particularly 
expensive to acquire. This paper contributes to the state-of-the-art with the following findings:
\begin{itemize}
\itemsep0em
\item A post-ASR text-to-text mapping can effectively adapt an
existing ASR system to a new language, providing a novel,
lightweight and low-cost strategy for ASR bootstrapping into large number of languages.
\item The text-to-text mapping is complementary with other bootstrapping
approaches, and it can improve an ASR system already bootstrapped 
with multilingual training and transfer learning.
\item Using audio generated by a text-to-speech (TTS) system alone is not 
sufficient to bootstrap ASR models, but it can be mixed with a small 
amount of real audio data to improve ASR performance further.
\end{itemize}

\section{Related Work}
\subsection{Existing ASR bootstrapping approaches}
\label{subsec:related_work_existing_ASR_bootstrapping_approaches}
For ASR bootstrapping, there are several approaches
in existing studies that leverage the benefit of data from data-rich languages
to low resource languages in ASR model training. One approach is 
\textit{transfer learning} \cite{kunze2017transfer,cho2018multilingual},
where the ASR model training for the target language is initialized using
parameters from an ASR model trained for other languages
(sometimes referred to as \textit{fine-tuning}).
Another approach is \textit{multilingual training}, i.e. train an ASR model jointly on 
available data from multiple languages, which was found particularly beneficial 
for low resource languages in previous study \cite{cho2018multilingual,pratap2020massively}.
Additionally, there is ongoing work on finding language agnostic
representations that enable rapid ASR expansion to new languages and tasks \cite{shor2020towards}.

\subsection{Post-ASR model}
\label{subsec:related_work_post-asr_model}
Previous studies have shown success in improving ASR performance with a post-ASR 
model.
Recent research on a post-ASR model has focused on applying it for ASR adaptation 
to a new domain \cite{mani2020asr} or
correcting frequent ASR errors \cite{d2016automatic, Guo2019}.
This work expands upon existing research on post-ASR models by 
applying it to a novel application context, i.e. ASR bootstrapping. 
In this context, a post-ASR model maps the incorrect recognition by an ASR system 
from a mismatched language to the desired transcription in the target language,
as well as corrects ASR mistakes made in the same language.

\subsection{Employ synthetic data from TTS to aid ASR}
\label{subsec:related_work_tts_for_asr}
Recent TTS research brought promising progress
to using TTS synthetic data for ASR model training,
especially when the data volume available for training is small.
A recent work \cite{Rosenberg2019} combined speech audio synthesized 
by TTS with real speech recordings, which improved the convergence of 
ASR model training when the available real data for training is as little as 10 hours. 
A follow-up work \cite{Rossenbach2020} showed that
adding synthetic data to E2E ASR model training is complementary
with SpecAugment \cite{Park2019SpecAugmentAS} 
and text based methods such as LM shallow fusion.
Previous research \cite{Rosenberg2019, Mimura2018} also demonstrated
the effectiveness of using TTS synthetic data to adapt an ASR model to a different domain.
In \cite{Zhang2021}, TTS synthetic data is found effective in teaching word piece based RNN-T
ASR models new vocabulary without the necessity of collecting additional real speech recordings.
This work extends existing research to a more extreme scenario with 
language mismatch, and uses TTS synthetic data to support ASR expansion 
to new languages without the time and financial cost from collecting a large quantity of audio recordings.

\section{Methodology}
Here, we combine three parallel threads from previous studies 
to bootstrap an RNN-T based ASR system 
for a new language, i.e. 1)
multilingual training and transfer learning;
2) employing a post-ASR text-to-text mapping;
3) using TTS synthetic audio to support ASR bootstrapping.

\subsection{RNN-T bootstrapping approaches}
\label{subsec:asr_bootstrapping_approaches}

This work covers two approaches to train RNN-T models: 
multilingual training \cite{kannan2019largescale} and transfer learning \cite{joshi2020transfer}. 
For both approaches, a multilingual RNN-T 
model is first trained on data from all four available non-target languages 
(English, German, Spanish and French), in addition to
a small amount of data from  the target language (Italian) when available. 
This approach extends previous work in \cite{joshi2020transfer}, and aims at obtaining a
better seed model for transfer learning.
When it comes to transfer learning, the trained multilingual RNN-T model is
fine-tuned towards the target Italian language
using the small amount of Italian data alone when it is available.
This approach is fundamental for an efficient usage of acoustic data available,
which is normally the most expensive to collect, compared to text data alone.

\subsection{Append a post-ASR text-to-text mapping to RNN-T}
\label{subsec:post_asr_t2t_model}
A post-ASR model is employed to map raw ASR output to desired 
transcriptions, thus a \textit{text-to-text mapping}.
Since one can apply it on top of any ASR system, it can be combined with all 
RNN-T bootstrapping approaches from 
Section \ref{subsec:asr_bootstrapping_approaches}. To build text-to-text mappings
for an existing ASR system and bootstrap its recognition performance on Italian, 
some Italian audio is decoded with the ASR system to produce paired text data.
Given a sample audio and its transcription, the paired text data consist of the raw ASR output hypothesis as the input 
for the text-to-text mapping, and the corresponding ground truth transcription as the targeted output
from the text-to-text mapping.
This approach is in spirit close to post-ASR error correction in \cite{Guo2019}.

The post-ASR text-to-text mappings used in this work are based on finite state 
transducer (FST, \cite{Bisani2008}), similar to 
the grapheme-to-phoneme mappings implemented by the Phonetisaurus package 
\cite{jiampojamarn2007applying,novak2016phonetisaurus}.
There are several considerations that motivated us to choose an FST model over
the more advanced neural models for ASR bootstrapping in this work.
First, an FST model is preferred over a neural model when operating in 
extremely low data regimes. 
Second, an FST model is easier to interpret and experiment with,
allowing a quick validation of the feasibility of this approach.
Third, the goal of a post-ASR text-to-text mapping is to convert 
the text representation of the same speech sound from one representation to another, 
with little complication of long-distance dependencies.
Therefore, it is fundamentally different from and easier than machine translation which 
heavily benefits from neural models.
This motivation was validated by a manual inspection into the input and 
output text pairs of an FST optimized to map the output of 
a multilingual seed RNN-T to Italian. An overwhelming
number of necessary mappings can be covered by simple rules involving local or 
limited context (examples in Table \ref{tab:fst_mapping_examples}). 

In this work, all FST models are built with word-level alignments, using N-gram FST 
with modified Kneser-Ney smoothing \cite{kneser1995improved}.
During training, the beam search decoding of the core ASR model produces a list of top 
hypotheses, which serves as the input for the N-gram FST models.
Using a relatively large beam size for ASR decoding ensures that the post-ASR text-to-text 
mapping model is trained on a rich variety of decoding hypotheses.
During inference, only the top 1 ASR hypothesis list is used as the input
for the N-gram FST models, thus a genuine post-ASR text-to-text mapping with minimal
additional computation cost introduced.
In our experiments we found that tuning the N-best list size in the FST decoding beam
and the N-gram context length for FST models had negligible impact on accuracy.
The results reported in this paper are based 5-gram FST models and a 500-best list configuration
is used for internal decoding of FST during both training and inference time.

\begin{table}[t]
  \caption{Example text-to-text mappings.  Left: multilingual seed RNN-T output in a mixture of English (EN), Spanish (ES), German (DE) and French (FR). Right: text-to-text mapping output and groundtruth in Italian (IT).}
  \vspace*{-0.3cm}
  \label{tab:fst_mapping_examples}
  \centering
  \begin{tabular}{ l l }
    \toprule
    \textbf{Multilingual ASR output} & \textbf{Desired truth (IT)} \\
    \midrule
    recommence (EN) & ricomincia \\
    she / c / see (EN) & sì \\
    repeating (EN) & ripeti \\
    i want tea (EN) & avanti \\
    cause of sci-fi (EN) & cosa sai fare \\
    bueno no te (ES) & buonanotte \\
    bonjour (FR) & buongiorno \\
    \bottomrule
  \end{tabular}
  \vspace*{-0.3cm}
\end{table}

\subsection{Generate synthetic data for ASR bootstrapping}
\label{subsec:tts_for_asr}
This work employs the Italian standard voices from Amazon Polly\footnote{https://aws.amazon.com/polly/}
to generate synthetic audio as an alternative data source for ASR bootstrapping
that is cheaper and faster to obtain than real labelled speech recordings. 
A few approaches are used to mitigate the risk of the acoustic mismatch between 
synthetic audio and real speech recordings. First, clean synthetic audio is corrupted with background noise and 
reverberation to increase the robustness of trained ASR model in far-field conditions.
Second, features of TTS synthetic data are normalized independently from 
features of real data. 
In addition, SpecAugment \cite{Park2019SpecAugmentAS} is applied on
all training data. Besides the widely observed regularization benefit, SpecAugment
also reduces the risk of overfitting an ASR model towards unrealistic synthetic data
and improves the ASR model robustness against challenging acoustic conditions
\cite{Rossenbach2020}.

\section{Experiments and Results}
\subsection{ASR system configuration}

ASR systems in this work are based on RNN-T \cite{rnnt}, with 
a transcription network (or encoder) of 5 LSTM layers (1024$\times$5), 
a prediction network (or decoder) of 2 LSTM layers (1024$\times$2), and
a joint network of one feed-forward layer.
The output units of RNN-T consist of 4000 wordpieces based on a unigram model trained 
with SentencePiece \cite{kudo2018sentencepiece} on 
multilingual text data from 13.4kh de-identified speech,
uniformly sampled from English, Spanish, German, French as well as Italian.

The acoustic features used are 64 dimensional log Mel filter banks
with a frame shift of 10ms. Similar to \cite{Pundak2016LFR}, the feature vectors 
from three consecutive frames are stacked together, leading to 
192 dimensional low frame rate acoustic features.
Global mean and variance feature normalization is applied per language, 
and independently for real data and TTS synthetic data. 
SpecAugment is further applied on-the-fly during training.
Two frequency masks are applied to each utterance and the maximal masked 
frequency percentage is 37.5\%. 
No time masks are used in our experiments. 
Where a mask applies, the actual feature value is replaced with a random value 
sampled from a Gaussian distribution following
the same mean and variance with the masked original feature values. 

ASR decoding is based on beam search with a beam size of 25 in all experiments and 
hypothesis cost is normalized by utterance length.
Besides providing a variety of ASR hypotheses to train text-to-text mappings,
a relatively large beam size is found to improve the robustness 
against the high deletion error rate on Italian audio caused by the complete absence or 
the insufficiency of real Italian training data.

Evaluation is based on word error rate (WER).
All results in this work are reported in normalized WER (NWER), i.e.
the absolute WER divided by a reference WER
which is a fixed value shared globally in this work. The reference WER
is from a monolingual Italian RNN-T trained on 160h real Italian data and
tested on 62h real Italian data (Model 1 in Table \ref{tab:multilingual} and Table \ref{tab:tts_on_low_resource}).
In addition, this 62h real Italian test data is used in all evaluations of follow-up
sections unless stated otherwise.

\vspace*{-0.1cm}
\subsection{Multilingual seed RNN-T}
\label{subsec:results_multilingual_seed_RNN-T}
\vspace*{-0.1cm}

As a baseline, a multilingual seed RNN-T (Model 2 in Table \ref{tab:multilingual}) 
is trained on data from four data-rich languages (English, German, Spanish 
and French). The amount of training data per language is shown in 
the ``Data (hour) - train" column of Table \ref{tab:multilingual},
along with the performance of this multilingual seed RNN-T 
on the test datasets from all four languages, in addition to a test dataset based on 
real Italian data even though the model has not seen any Italian data during training.

The first observation is that the multilingual RNN-T trained without any Italian data
is still capable of recognizing a small proportion of Italian utterances.
This is because some voice commands are fully or partly
shared across languages. For example, the wake-word used to
trigger voice assistants is shared across languages, so are the popular
voice control phrases such as \textit{stop}, \textit{playlist} and \textit{to-do list}.
Furthermore, the multilingual model also correctly recognizes international entity
names.
For example, the music group \textit{Rammstein} is correctly recognized in Italian
because it also appears in German data.

%

\begin{table}[t]
  \caption{The performance of Italian monolingual RNN-T (Model 1) on real Italian data, and multilingual seed RNN-T (Model 2) on test data from different languages. (R: RNN-T trained on real Italian data only; M: RNN-T trained on multilingual data from 4 foreign languages excluding Italian.)}
  \vspace*{-0.3cm}
  \label{tab:multilingual}
  \centering
  \begin{tabular}{ l| c| l r r r r r r }
    \toprule
    \textbf{Model} & \textbf{RNN-T} & \multirow{2}{*}{\textbf{Language}} & \multicolumn{2}{c}{\textbf{Data (hour)}} &  \textbf{NWER} \\
    \textbf{index} & \textbf{seed} &  & \textbf{train} & \textbf{test} & (\%) \\
    \midrule
    1 & R 	& Italian 		& 160 			& 62 	& \textbf{1.00} \\  
    \midrule
    2 & M 	& German  	& 13800 		& 72		& 0.48 \\
    2 & M 	& English 		& 39800 		& 169 	& 0.38 \\
    2 & M 	& Spanish 	& 7100  		& 75 		& 0.47 \\
    2 & M 	& French  		& 6700  		& 72 		& 0.55 \\
    2 & M 	& Italian 		& -     			& 62 	& \textbf{2.76}	\\
    \bottomrule
  \end{tabular}
\end{table}

\vspace*{-0.1cm}
\subsection{Bootstrap with multilingual training, transfer learning and text-to-text mapping}
\label{subsec:results_different_bootstrapping_methods}
\vspace*{-0.1cm}

\begin{figure}[t]
  \centering
  \includegraphics[width=\linewidth, trim=0 0.5cm 0 0.4cm, clip]{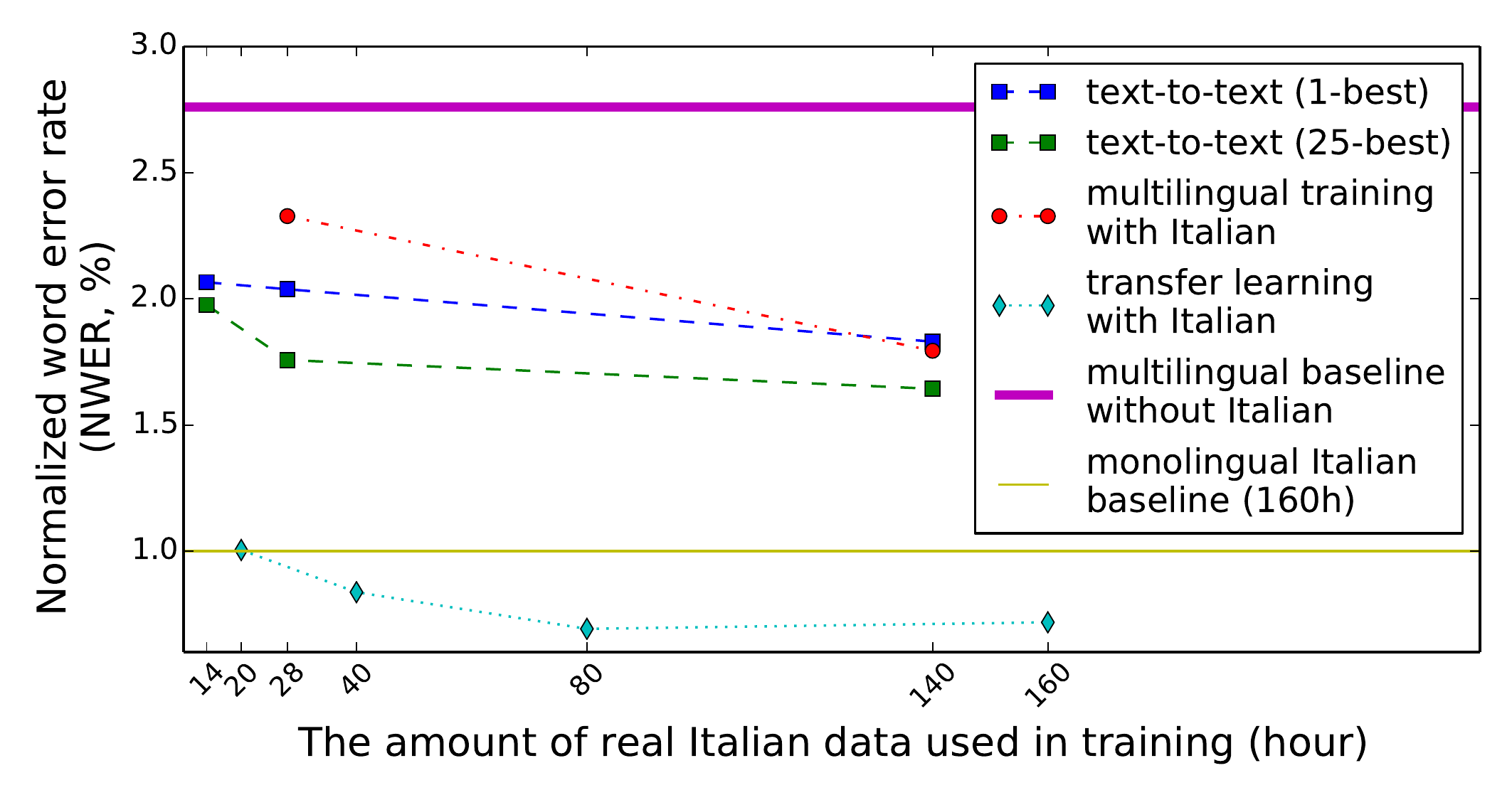}
  \caption{
  NWER of different bootstrapping approaches, plotted as a function of the hours of real Italian audio ingested.
  }
  \vspace*{-0.5cm}
  \label{fig:t2t_vs_direct}
\end{figure}

Different bootstrapping approaches are compared on the same real Italian test data.
In addition, the performance is examined given different amounts of available
real Italian training data. 

In Figure \ref{fig:t2t_vs_direct}, the ``multilingual baseline without Italian'' refers to the 
multilingual seed RNN-T from Section \ref{subsec:results_multilingual_seed_RNN-T}.
The post-ASR text-to-text mapping is added on top of this multilingual seed RNN-T,
in two configurations that takes the 1-best list and the 25-best list 
of ASR hypothesis for text-to-text mapping training respectively (``text-to-text'' in Figure \ref{fig:t2t_vs_direct}).
As shown, both configurations decreased NWER by a large margin compared to the multilingual baseline.
In addition, given different quantities of available Italian training data, 
using 25-best ASR output to train the post-ASR model consistently outperforms using 
1-best ASR output , with a relative improvement of approximately 5\% to 15\%.
In a dive deep analysis, we found that the multilingual seed RNN-T makes a lot of deletion 
errors on the test data of the unseen Italian language, leading to nonsensical text-to-text mappings. 
This issue is then partially mitigated by using a large N-best list.

The available Italian data is also added directly to multilingual training 
(``multilingual training with Italian'' in Figure \ref{fig:t2t_vs_direct}).
As shown, adding 140h Italian data directly to multilingual training gives performance worse than but 
close compared to using a text-to-text mapping (``text-to-text'' in Figure \ref{fig:t2t_vs_direct}),
while the latter has a clear advantage at the extremely-low data regimes (28h).

Overall, directly adapting the ``multilingual training with Italian" model with available Italian data 
consistently tops among all investigated approaches
(``transfer learning with Italian" in Figure \ref{fig:t2t_vs_direct}), even though the final model performance
certainly depends on the similarity between the newly bootstrapped language and the ones used to
build the seed model.

\subsection{Use synthetic audio for ASR bootstrapping}

To simulate the scenario where only in-domain text is available from the
target language, the transcription of real Italian voice assistant
training data is used to generate TTS audio. In our experiments 0.7 million
in-domain text utterances are used for TTS to synthesize training data.
Obtaining this amount of in-domain data is not obviously easy, but techniques like machine translation
and named entities resampling can help.
Each text utterance is randomly paired with one of 3 available Italian voice profiles 
from Amazon Polly
standard voice (1 male voice and two female voices), making 280h TTS audio in total. 
In a previous benchmark \cite{Prateek2019}, 
the Polly standard voice offered at least an 80\% relative MUSHRA and MOS scores in naturalness.

Since TTS synthetic audio is artificially clean compared to real speech recordings,
using it directly to train ASR models poses risks on
the acoustic robustness when the trained models are evaluated on far-field recordings. 
Therefore, the 280h TTS audio is combined with
a copy where the synthetic audio is corrupted with non-speech noise
and reverberation, this makes 560h Italian TTS synthetic data in total for training.
To produce the corrupted copy,  each TTS synthetic audio utterance is convolved with 
an acoustic impulse response (AIR) to simulate reverberant speech. 
The AIR is randomly selected from a 10k pool of AIR data collected via chirp signal based measurement
in real rooms using real Echo devices.
On top of that, a randomly selected segment of non-speech noise
audio recording is mixed with the reverberated audio to simulate far-field recording with background noise.
The noise recordings covers typical domestic noise such as kitchen noise and vacuum noise,
as well as synthetic noise such as pink noise and white noise.
The signal-to-noise ratio (SNR) between the reverberated speech audio and the 
background noise follows a Gaussian distribution,
with an average SNR of 20dB and a standard deviation of 8dB.

The second section of Table \ref{tab:tts_on_low_resource} shows
the ASR performance with the adoption of TTS synthetic when there is
no real Italian speech recordings used for training.
In particular, Model 4 uses Italian synthetic data to optimize the
text-to-text mapping, while Model 5 combines Italian synthetic data with
non-Italian multilingual real data for multilingual RNN-T training.
Compared to the multilingual RNN-T baseline (Model 2),
both Model 4 and Model 5 reduced the NWER on real Italian test data,
though the improvement is modest. 
To gain insights into the modest improvement, an Italian TTS testset was generated without audio corruption.
When tested on this TTS testset, Model 5 achieved an NWER of 0.27. 
This is much lower than the NWER on the counterpart real Italian test set based on human speech recordings (2.41). 
This indicates that such a model has problems generalizing to real speech recordings from the synthetic speech
audio seen in training.
When looking into recognition errors, we observed that Model 5 properly decoded Italian TTS utterances into Italian language,
while real human Italian speech was decoded either as speech from other languages or empty output. 
Despite the data augmentation efforts aimed at reducing the acoustic gap between
clean synthetic audio and far-field speech recordings, the models optimized with TTS 
synthetic data alone in the targeted language still faces challenges when it comes to 
recognizing real speech recordings.

This motivates us to mix TTS synthetic audio and real recordings for RNN-T 
transfer learning when real speech recordings in the target Italian language are available.
The third section of Table \ref{tab:tts_on_low_resource} shows the results of
such experiments, where 560h synthetic data are mixed with 160h real data in model optimization.
Consistent with previous observations, adding a post-ASR text-to-text mapping
consistently outperforms all corresponding baseline ASR systems
(Model 6 vs. Model 2, Model 8 vs. Model 7 and Model 10 vs. Model 9),
and the improvement is larger at poorer performance baseline
(Model 6 vs. Model 2). 
Comparing Model 9 to Model 7, adding TTS synthetic audio to real data for
RNN-T transfer learning improved recognition performance by 22\% relative.
Furthermore, transfer learning and text-to-text mapping combine well to bring cumulative
accuracy improvements (Model 10 compared to Models 7, 8 and 9).
The best performance is achieved with Model 10 that combines multilingual
training, transfer learning, post-ASR text-to-text mappings
and TTS synthetic data. Model 10 is 46\% relatively better compared to the
monolingual baseline (Model 1), 
and 25\% relatively ahead of Model 7 where neither
using text-to-text mappings nor any synthetic data is used.
Moreover, comparing the results in Table \ref{tab:multilingual} and Table \ref{tab:tts_on_low_resource},
the performance of Model 10 with just 160h of real Italian data
is approaching to the performance of the multilingual model on other languages
trained on one or two orders of magnitude more data.


\begin{table}[t]
  \caption{
  Use 560h Italian TTS synthetic audio for Italian ASR bootstrapping,
  optionally mixed with 160h real Italian data.
  All text-to-text mapppings have been trained with 25-best hypotheses.
  (M: multilingual data excluding Italian; R: real Italian data;
  TTS: Italian TTS synthetic audio)
  }
  \label{tab:tts_on_low_resource}
  \centering
  \begin{tabular}{ l| l| l l r r r }
    \toprule
     \textbf{Model} & \textbf{RNN-T} & \textbf{Transfer} & \multirow{2}{*}{\textbf{Text-to-text}} & \textbf{NWER} \\
     \textbf{index} & \textbf{seed} &  \textbf{learning} &  & (\%) \\
    \midrule
    1 & R & - & - & 1.00 \\ 
    2 & M & - & - & 2.76 \\ 
    3 & M + R & - & - &  1.80\\ 
    \midrule
    4 & M & - & TTS & 2.43 \\
    5 & M + TTS & - & - & 2.41 \\
    \midrule
    6 & M & - & R + TTS & 1.61 \\
    7 & M + R & R & - & 0.72 \\
    8 & M + R & R & R + TTS & 0.65 \\
    9 & M + R & R + TTS & -  & 0.56 \\
    10 & M + R & R + TTS & R + TTS & \textbf{0.54} \\
    \bottomrule
  \end{tabular}
\end{table}

\section{Conclusions}
In a simulated Italian ASR bootstrapping scenario, this work demonstrated 
that using a post-ASR text-to-text mapping and synthetic data generated
with TTS brings extra benefit on top of RNN-T bootstrapping approaches such as
multilingual training and transfer learning. The best ASR system employed
all methods in combination, achieving a performance 46\% relatively better than a monolingual
baseline, and 25\% relatively better than not using post-ASR text-to-text
mapping or any synthetic data.
We note that the bootstrapped model achieves a performance comparable
to the one achieved by the multilingual model on the languages with much larger volumes of data.
We further note that synthetic audio from TTS was of limited benefit
when used on its own, but it became beneficial for
RNN-T transfer learning and post-ASR text-to-text mappings
when it is combined with small amounts of human audio.

Future work will investigate the benefit of using word-piece or all-neural 
models for post-ASR text-to-text mapping. 
In addition, phonetic units will be explored as to providing complementary
information for text-to-text mapping.

\bibliographystyle{IEEEtran}
\bibliography{bibliography}

\end{document}